\documentclass[prl,twocolumn,superscriptaddress]{revtex4}
\usepackage{graphicx,color}
\usepackage{amssymb}   % for math
\usepackage{amsmath}
\usepackage{epstopdf}
\usepackage{natbib}
\usepackage{hyperref}
\usepackage{bm}
\usepackage{color}
\usepackage{tabularx}
\usepackage{multirow}
\usepackage{braket}
\usepackage{xcolor}

\begin{document}

\title{Kramers Weyl Semimetals as Quantum Solenoids and Their Applications in Spin-Orbit Torque Devices}
\author{Wen-Yu He}\thanks{wenyuhe@mit.edu}
\affiliation{Department of Physics, Massachusetts Institute of Technology, Cambridge, Massachusetts 02139, USA}
\author{Xiao Yan Xu}
\affiliation{Department of Physics, University of California at San Diego, La Jolla, California 92093, USA}
\author{K. T. Law}\thanks{ phlaw@ust.hk}
\affiliation{Department of Physics, Hong Kong University of Science and Technology, Clear Water Bay, Hong Kong, China}
\date{\today}

\begin{abstract}
{\emph{Abstract}---Kramers Weyl semimetals are Weyl semimetals that have Weyl points pinned at the time reversal invariant momenta. Recently it has been discovered that all chiral crystals host Weyl points at time reversal invariant momenta, so metals with chiral lattice symmetry all belong to the category of Kramers Weyl semimetals. In this work, we show that due to the chiral lattice symmetry, Kramers Weyl semimetals have the unique longitudinal magnetoelectric effect in which the charge current induced spin and orbital magnetization is parallel to the direction of the current. This feature allows Kramers Weyl semimetals to act as nanoscale quantum solenoids with both orbital and spin magnetization. As the moving electrons of Kramers Weyl semimetal can generate longitudinal magnetization, Kramers Weyl semimetals can be used for new designs of spin-orbit torque devices with all electric control of magnetization switching for magnets with perpendicular magnetic anisotropy.}
\end{abstract}

\maketitle

\section{Introduction}
Weyl semimetals are nodal topological materials characterised by isolated band touching points, called Weyl points, in 3D momentum space~\cite{Xiangang, Balents1, Balents2, Weng}. Due to the Weyl points, which act as the monopoles of the Berry curvature in the momentum space~\cite{Xiangang, Balents1, Balents2, Weng}, Weyl semimetals exhibit many exotic properties such as the chiral magnetic effect~\cite{Ninomiya, Yamamoto, Genfu, Chenglong}, the presence of topologically protected Fermi arcs states~\cite{Xiangang, Balents1, Balents2, Weng, Suyang, Lv}, unconventional quantum oscillations~\cite{Potter}, and novel optical phenomenon~\cite{Chris1, Chris2, Moore0}. Recently, a new type of Weyl semimetals called Kramers Weyl semimetals (KWS) in chiral crystals have been discovered~\cite{Bradlyn, Shoucheng, Guoqing2, Guoqing, Hasan, DingHong, Yulin, Ando, Felser1, Felser2}.  Chiral crystals are crystals which lack inversion, mirror and improper rotation symmetries. As a result, a chiral crystal has a definite handedness and can be described by the 11 chiral point groups. It was shown that generally, due to the low lattice symmetry, band splittings appear away from time-reversal invariant points in momentum space and result in Kramers Weyl points pinned at time-reversal invariant momenta. However, it is not clear how the properties of KWS are distinct from that of other Weyl semimetals with non-chiral point group symmetry. 

In this work, we point out that the chiral lattice symmetry in KWS brings about a unique property: an electric field applied along the principal symmetry axis of the crystal would induce spin and orbital magnetization which is parallel to the applied electric field. This is in sharp contrast to the case of all other noncentrosymmetric Weyl semimetals which give \emph{zero} magnetoelectric response if the electric field is applied along the principal symmetry axis. This distinctive longitudinal magnetoelectric response involves both the spin~\cite{Edelstein} and orbital magnetization~\cite{Moore, Jing}, and it arises from the special form of the spin-orbit coupling (SOC) of KWS. Importantly, for the representative KWS K$_2$Sn$_2$O$_3$ and RhSn in the T point group we considered, the induced magnetization at a given electric field can be two to three orders of magnitude larger than the magnetization induced in materials with the strong Rashba spin-orbit couplings such as in Au (111) surfaces and Bi/Ag bilayers~\cite{Mertig1, Mertig2}. Therefore, if the electrons which carry both orbital and spin angular momentum in the KWS are injected into a ferromagnetic layer, the torque induced by the electrons can cause magnetization switching in the ferromagnetic layer. As a result, we propose a KWS based magnetization switching device that is different from the magnetic tunneling device based on spin transfer torque~\cite{Manchon, Sinova} and the conventional spin orbit torque devices~\cite{Manchon, Sinova, Ralph, Jungwirth}. The new KWS based devices allow all electric control of magnetic switching for ferromagnets with perpendicular magnetic anisotropy, which is important for high density magnetic memories.

\section{Results}
\subsection{Effective Hamiltonians for Kramers Weyl Semimetals}
In chiral crystals which respect time reversal symmetry, the energy bands are at least doubly degenerate at time-reversal invariant momenta due to the Kramers theorem. In the absence of inversion, mirror and improper rotation symmetries in chiral crystals, and away from the time-reversal invariant points, SOC would lift the Kramers degeneracy in momentum space to create Kramers Weyl points~\cite{Guoqing}. To be more specific, in the spin $\frac{1}{2}$ basis $\psi_{\bm{k}}=\left[\phi_{\bm{k}, \uparrow}, \phi_{\bm{k}, \downarrow}\right]^{\textrm{T}}$, which satisfies the relation $\Theta\psi_{\bm{k}}=i\sigma_y\psi_{-\bm{k}}$ under time-reversal operation $\Theta=i\sigma_yK$, the effective Hamiltonian $H_0\left(\bm{k}\right)$ can be obtained through standard $\bm{k}\cdot\bm{p}$ method~\cite{Guoqing}. Up to second order near a time-reversal invariant momentum $\bm{k}_0$, the general form of the Kramers Weyl Hamiltonian can be written as
\begin{align}
H_0\left(\bm{k}\right)=\sum_{i, j}\frac{\hbar^2}{2m_{ij}} k_i k_j+\sigma_i\hbar v_{ij} k_j.
\end{align}
Here, $\bm{k}$ is measured from $\bm{k}_0$, $i, j=x, y, z$, $m_{ij}$ is the effective mass tensor, $\sigma_i$ are the Pauli matrices in spin space, and $v_{ij}$  is the SOC pseudotensor. In chiral crystals, the little group at $\bm{k}_0$ is isomorphic to a chiral point group which guarantees $\det\left(v\right)\neq0$ so that the Kramers Weyl point emerges at $\bm{k}=0$.

In the KWS, the specific forms of the SOC that creates the Weyl point are determined by the point group symmetry as $v=\det\left(\hat{R}\right)\hat{R}v\hat{R}^{\textrm{T}}$, with $\hat{R}$ the symmetry transformation matrix. In materials within the cubic point group $\left\{\textrm{T}, \textrm{O}\right\}$ such as K$_2$Sn$_2$O$_3$~\cite{Guoqing}, RhSi~\cite{Hasan}, CoSi~\cite{Hasan, DingHong, Ando}, AlPt~\cite{Yulin}, PtGa~\cite{Felser1}, and PdGa~\cite{Felser2} of cubic chiral B20 structure~\cite{Burkov}, at $\bm{k}_0$ (such as the $\Gamma$ point) the little group isomorphic to $\left\{\textrm{T}, \textrm{O}\right\}$ can give rise to the isotropic Weyl Hamiltonian
\begin{align}\label{isotropic}
H_0\left(\bm{k}\right)=\frac{\hbar^2}{2m}\bm{k}^2+\hbar v\bm{k}\cdot\bm{\sigma},
\end{align}
where the high symmetry cubic point group $\left\{\textrm{T}, \textrm{O}\right\}$ forces $v_{ij}$ to be proportional to the identity matrix. In those materials within the point group $\left\{\textrm{T}, \textrm{O}\right\}$, the high symmetry may also enable extra multi-fold point degeneracies to appear at $\bm{k}_0$~\cite{Bradlyn, Shoucheng, Guoqing2}. In the materials CsCuBr$_3$~\cite{Guoqing}, elemental Te, Se~\cite{Miyake,Itou, Takeshi, Tetsuaki}, etc., the dihedral point group there has lower crystal symmetry and allows the anisotropy to show up in the Weyl Hamiltonian
\begin{align}
H_0\left(\bm{k}\right)=\sum_{i=x, y, z}\frac{\hbar^2}{2m_i}k_i^2+\hbar v_i\sigma_ik_i.
\end{align}
In the materials belonging to the cyclic point group, as the crystal symmetry is further reduced, the constraints on the SOC is further reduced. The complete forms of the SOC pseudotensor $v_{ij}$ in the KWS Hamiltonian are summarised in the Supplementary Table 1 for all the chiral point groups.

The SOC in the KWS creates the Kramers Weyl points at the time reversal invariant momenta and allows the coupling between the spin and momentum. Under an electric field, the SOC enables the charge carriers to have net magnetization and such magnetoelectric response respects the same crystal symmetry present in the SOC. As shown below, for materials with cubic point groups $\left\{\textrm{T}, \textrm{O}\right\}$, the simple form of the isotropic Weyl Hamiltonian in Eq. \ref{isotropic} allows us to calculate the magnetoelectric susceptibility analytically. For the KWS in dihedral and cyclic point group, the magnetoelectric responses are calculated numerically for the selected materials in Supplementary Note 1 and Supplementary Note 2 respectively.

\subsection{Magnetoelectric Pseudotensors and Their Symmetry Properties}
In magnetoelectric effects, induced magnetization $\bm{M}$ and the applied electric field $\bm{E}$ are related by the magnetoelectric pseudotensor $\alpha$ such that:
\begin{align}
M_i=\sum_{i, j}\alpha_{ij}E_j,
\end{align}
where $i, j=x, y, z$ and $\alpha_{ij}$ are elements of the magnetoelectric pseudotensor $\alpha$. 
For a generic Hamiltonian
\begin{align}
\mathcal{H}=\sum_{\nu, \nu', \bm{k}}c^\dagger_{\nu, \bm{k}}H_{0, \nu\nu'}\left(\bm{k}\right)c_{\nu', \bm{k}},
\end{align}
where $c^\dagger_{\nu, \bm{k}}\left(c_{\nu, \bm{k}}\right)$ is the creation (annihilation) operator, $H_{0, \nu\nu'}\left(\bm{k}\right)$ is the element of the Hamiltonian matrix $H_0\left(\bm{k}\right)$,  $\alpha_{ij}$ can be obtained from the linear response theory as~\cite{Moore, Jing}
\begin{align}\label{ME_sus}
\alpha_{ij}=-\tau\frac{e}{\hbar}\frac{1}{\left(2\pi\right)^d}\int_{\textrm{BZ}}d\bm{k}\sum_n M_{n\bm{k}, i}v_{n\bm{k}, j}\frac{df\left(E_{n\bm{k}}\right)}{dE_{n\bm{k}}}.
\end{align}
In Eq.\ref{ME_sus}, $f\left(E_{n\bm{k}}\right)$ is the Fermi Dirac distribution function, $E_{n\bm{k}}$ is the energy dispersion of band $n$ from the Hamiltonian $H_0\left(\bm{k}\right)$, $v_{n\bm{k}, j}=\frac{\partial E_{n\bm{k}}}{\partial k_j}$, $d$ is the dimension of the system, $\tau$ is the effective scattering time and $i, j=x, y, z$ denote the spatial components. The total magnetic moment $\bm{M}_{n\bm{k}}=\bm{S}_{n\bm{k}}+\bm{m}_{n\bm{k}}$ carried by the Bloch electrons consists of both the spin magnetic moment $\bm{S}_{n\bm{k}}=\bra{\phi_{\bm{k}, n}}\frac{1}{2}g\mu_{\textrm{b}}\bm{\sigma}\ket{\phi_{\bm{k}, n}}$ and the orbital magnetic moment $\bm{m}_{n\bm{k}}=\frac{ie}{2\hbar}\bra{\partial_{\bm{k}}\phi_{\bm{k}, n}}\times\left[H_0\left(\bm{k}\right)-E_{n\bm{k}}\right]\ket{\partial_{\bm{k}}\phi_{\bm{k}, n}}$. Here, $\mu_{\textrm{b}}=\frac{e\hbar}{2m_{\textrm{e}}}$ is the Bohr magneton, $g$ is the Lande $g$ factor which is set to be 2 in our calculations and $\ket{\phi_{\bm{k}, n}}$ denotes a Bloch state. As we will show explicitly below, the orbital magnetization is related to the Berry curvature of the Bloch states which has the form $\bm{\Omega}_{n\bm{k}}=i\bra{\partial_{\bm{k}}\phi_{\bm{k}, n}}\times\ket{\partial_{\bm{k}}\phi_{\bm{k}, n}}$ ~\cite{Niuqian}.

\begin{table}[ht]
\caption{of Magnetoelectric susceptibility pseudotensor $\alpha$ for the chiral crystals in the 11 chiral point groups. $\alpha_{ij}$ with $i, j=x, y, z$ are in general the elements in $\alpha$. In point group with symmetry, the 9 elements in $\alpha_{ij}$ is no longer independent. $\alpha_0$ means $\alpha_0=\alpha_{xx}=\alpha_{yy}=\alpha_{zz}$ in T and O point group. In the \{C$_3$, C$_4$, C$_6$, D$_3$, D$_4$, D$_6$\} group $\alpha_{xx}=\alpha_{yy}$ is then denoted as $\alpha_{\parallel}=\alpha_{xx}=\alpha_{yy}$. $\alpha^-$ means the antisymmetric elements as $\alpha^-=-\alpha_{xy}=\alpha_{yx}$ in group \{C$_{3}$, C$_4$, C$_6$\}. The principal axis of the crystal is set along $z$.} % title of Table
\centering % used for centering table
\begin{tabular}{c c c c c c} % centered columns (4 columns)
\hline\hline %inserts double horizontal lines
Point group & $\alpha$ & Point group & $\alpha$\\ [0.5ex] % inserts table
%heading
\hline % inserts single horizontal line
O & $\begin{pmatrix}
\alpha_0 & 0 & 0 \\ 
0 & \alpha_0 & 0 \\ 
0 & 0 & \alpha_0
\end{pmatrix}$ & T & $\begin{pmatrix}
\alpha_0 & 0 & 0 \\ 
0 & \alpha_0 & 0 \\ 
0 & 0 & \alpha_0
\end{pmatrix}$ \\
D$_{2}$ & $\begin{pmatrix}
\alpha_{xx} & 0 & 0 \\ 
0 & \alpha_{yy} & 0 \\ 
0 & 0 & \alpha_{zz}
\end{pmatrix}$ & D$_3$ & $\begin{pmatrix}
\alpha_{\parallel} & 0 & 0 \\ 
0 & \alpha_{\parallel} & 0 \\ 
0 & 0 & \alpha_{zz}
\end{pmatrix}$ \\
D$_4$ & $\begin{pmatrix}
\alpha_{\parallel} & 0 & 0 \\ 
0 & \alpha_{\parallel} & 0 \\ 
0 & 0 & \alpha_{zz}
\end{pmatrix}$ & D$_6$ & $\begin{pmatrix}
\alpha_{\parallel} & 0 & 0 \\ 
0 & \alpha_{\parallel} & 0 \\ 
0 & 0 & \alpha_{zz}
\end{pmatrix}$ \\C$_1$ & $\begin{pmatrix}
\alpha_{xx} & \alpha_{xy} & \alpha_{xz} \\ 
\alpha_{yx} & \alpha_{yy} & \alpha_{yz} \\ 
\alpha_{zx} & \alpha_{zy} & \alpha_{zz}
\end{pmatrix}$ & C$_2$ & $\begin{pmatrix}
\alpha_{xx} & \alpha_{xy} & 0 \\ 
\alpha_{yx} & \alpha_{yy} & 0 \\ 
0 & 0 & \alpha_{zz}
\end{pmatrix}$ \\
C$_3$ & $\begin{pmatrix}
\alpha_{\parallel} & -\alpha^- & 0 \\ 
\alpha^- & \alpha_{\parallel} & 0 \\ 
0 & 0 & \alpha_{zz}
\end{pmatrix}$ & C$_4$ & $\begin{pmatrix}
\alpha_{\parallel} & -\alpha^- & 0 \\ 
\alpha^- & \alpha_{\parallel} & 0 \\ 
0 & 0 & \alpha_{zz}
\end{pmatrix}$ \\ % inserting body of the table
C$_6$ & $\begin{pmatrix}
\alpha_{\parallel} & -\alpha^- & 0 \\ 
\alpha^- & \alpha_{\parallel} & 0 \\ 
0 & 0 & \alpha_{zz}
\end{pmatrix}$ &  \\[1ex] % [1ex] adds vertical space
\hline %inserts single line
\end{tabular}
\label{magnetoelectric_pseudotensor} % is used to refer this table in the text
\end{table}

The linear response theory applies to generic Hamiltonians. However, to shed light on the general properties of KWS, we note that the form of $\alpha$ can be determined by point group symmetries which is independent of the details of the Hamiltonian. The group theory analysis of $\alpha$ is elaborated in the Method Section as well as in Ref.~\cite{ Wenyu2}, and the general form for the chiral point groups is provided in Table \ref{magnetoelectric_pseudotensor}. From the group theory point of view, KWS can be classified into three sub-classes. For KWS belonging to the cubic point groups $\left\{\textrm{T}, \textrm{O}\right\}$ ,  $\alpha$ is proportional to the identity matrix as shown in Table \ref{magnetoelectric_pseudotensor}. This implies that the induced magnetization is always parallel to the direction of the applied electric field. Therefore, these KWS can behave as classical solenoids in all electric field directions without the need to fabricate any spiral structures.

For KWS with point groups D$_{n}$, a pure longitudinal magnetization parallel to the electric field is also obtained when the electric field is applied along the direction of any of the symmetry axes. For KWS with cyclic point groups, in general, magnetization with both components parallel and perpendicular to the direction of the applied electric field are generated. Interestingly, for all other Weyl semimetals without chiral point group symmetry, the magnetoelectric response is zero if the electric field is applied along the principal axis, which can be obtained from the Supplementary Table 1. Therefore, the longitudinal magnetoelectric response along the principal symmetry axis is a very distinctive feature of KWS due to the special spin texture of KWS which determines the spin and orbital magnetization.

Combined with the effective Hamiltonian for isotropic KWS in Eq. \ref{isotropic}, the magnetoelectric susceptibility in Eq. \ref{ME_sus} can be explicitly calculated and it shows how the SOC strength and band dispersion will influence the magnetoelectric response.

\subsection{Longitudinal Magnetoelectric Response in KWS}

\begin{figure}
\centering
\includegraphics[width=3.6in]{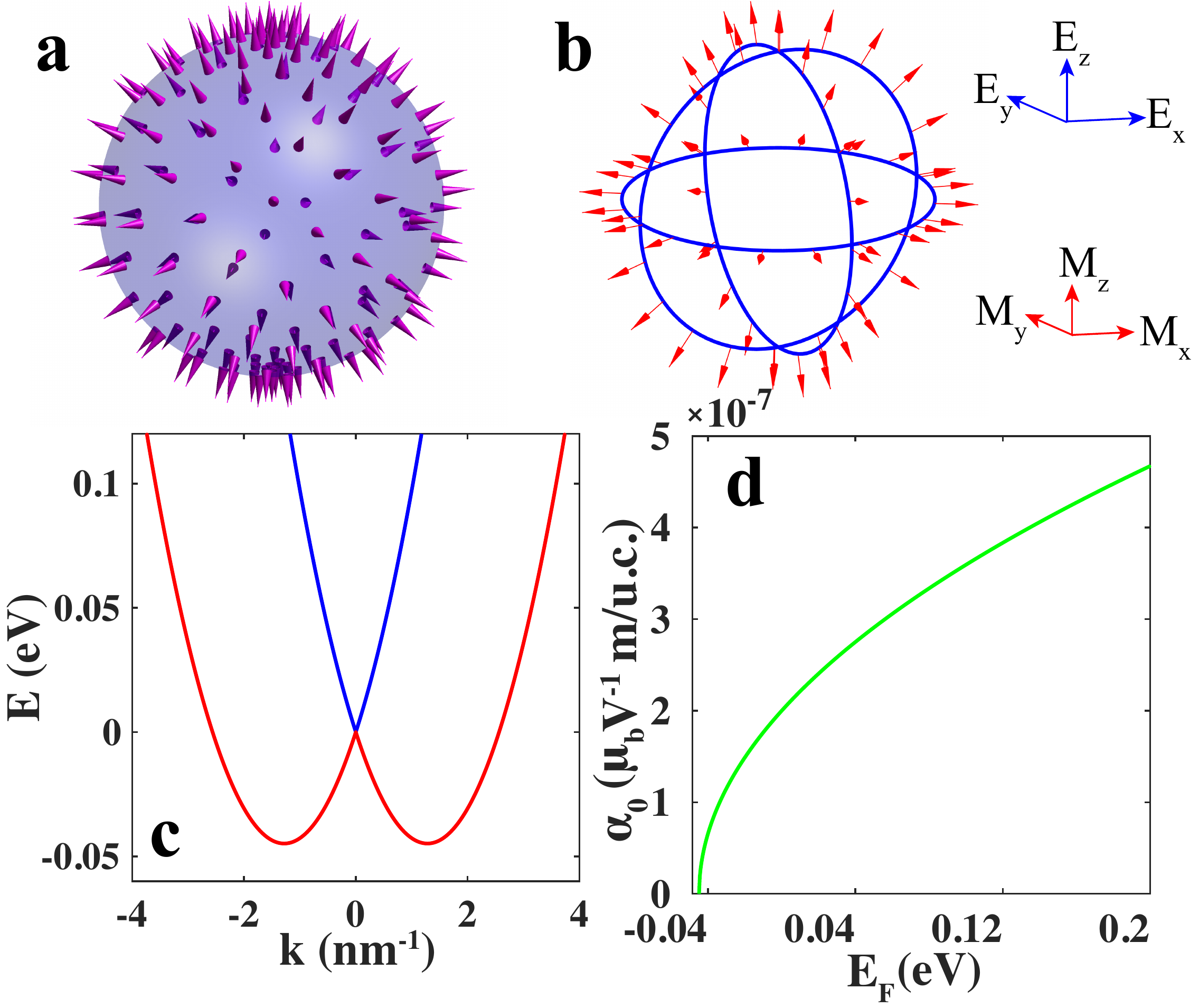}
\caption{The longitudinal magnetoelectric response of isotropic Kramers Weyl semimetals. (a) The Weyl spin texture of isotropic Weyl spin-orbit coupling $\hbar v\bm{k}\cdot\bm{\sigma}$ at the Fermi surface from the band branch $+$. (b) The electrically induced magnetization is parallel to the applied electric field. Here blue and red coordinates are the Cartesian coordinates for the applied electric field $\bm{E}=\left(E_x, E_y, E_z\right)$ and the generated magnetization $\bm{M}=\left(M_x, M_y, M_z\right)$. (c) The energy dispersion for the isotropic chiral Weyl semimetal. The blue and red line corresponds to the $+$ and $-$ band branch respectively. (d) The magnetoelectric susceptibility $\alpha_0$ strength as a function of the Fermi energy $E_{\textrm{F}}$. In the presence of external electric field, $\alpha_0$ gives the number of Bohr magneton $\left(\mu_{\textrm{b}}\right)$ per unit cell (u.c.) to denote the magnetization in the material.}\label{figure2}
\end{figure}

We first consider an effective Hamiltonian which describes a Kramers Weyl point near the $\Gamma$ point in chiral crystals with point groups $\left\{\textrm{T}, \textrm{O}\right\}$, where the isotropic Weyl Hamiltonian is $H_0\left(\bm{k}\right)=\frac{\hbar^2}{2m}\bm{k}^2+\hbar v\bm{k}\cdot\bm{\sigma}$. At the Fermi energy $E_{\textrm{F}}=\frac{\hbar^2}{2m}\bm{k}_{\textrm{F}}^2\pm \hbar v|\bm{k}_{\textrm{F}}|$, there are two spherical Fermi surfaces with corresponding wave vectors $\bm{k}_{\textrm{F}\pm}= k_{\textrm{F}\pm}\hat{\bm{k}}$, where $ k_{\textrm{F}\pm}=\frac{1}{\hbar}\sqrt{2mE_{\textrm{F}}+m^2v^2}\mp \frac{mv}{\hbar}$ and $\hat{\bm{k}}=\frac{\bm{k}}{|\bm{k}|}$. The spin and orbital magnetic moments of the two Fermi surfaces with Fermi momenta $\bm{k}_{\textrm{F}\pm}$ can be written as
\begin{align}\label{Magnetic_moment}
\bm{S}_{\bm{k}_{\textrm{F}}\pm}=\pm\frac{g}{2}\frac{e\hbar}{2m_{\textrm{e}}}\hat{\bm{k}}_{\textrm{F}\pm},\quad\textrm{and}\quad\bm{m}_{\bm{k}_{\textrm{F}}\pm}=\frac{ev}{2}\frac{\hat{\bm{k}}_{\textrm{F}\pm}}{|\bm{k}_{\textrm{F}\pm}|}.
\end{align}
It is important to note that the orbital magnetic moment $\bm{m}_{\bm{k}_{\textrm{F}}\pm}$ is proportional to the Berry curvature generated by the Weyl point inside Fermi surfaces which is $\bm{\Omega}_{\bm{k}_{\textrm{F}\pm}}=\mp\frac{\hat{\bm{k}}_{\textrm{F}\pm}}{2\bm{k}^2_{\textrm{F}}\pm}$. The spin texture on a Fermi surface is schematically shown in Fig.\ref{figure2}a.  It is clear that without breaking time-reversal symmetry, the total magnetic moment of all the electrons is zero. By applying an electric field, the steady state distribution of the electronic state can generate a net magnetization as indicated in Eq. \ref{ME_sus}. With this special form of spin texture of an isotropic KWS,  at the Fermi energy $E_{\textrm{F}}$, we obtain the isotropic longitudinal magnetoelectric susceptibility $\alpha_0$ as (see the method section for detailed derivation)
\begin{align}\label{ME_isotropic}
\alpha_0=-\frac{e^2v\tau}{6\hbar^2\pi^2}\sqrt{2mE_{\textrm{F}}+m^2v^2}\left(g\frac{m}{m_{\textrm{e}}}-1\right).
\end{align}
In $\alpha_0$, the first and second terms are the spin and the orbital contributions respectively. It is important to note that, for hole bands with negative effective mass, the spin and orbital contribution will always add together to enhance the magnetoelectric response. From Table \ref{magnetoelectric_pseudotensor},  the magnetoelectric response of an isotropic KWS can be written as $\bm{M}=\alpha_0 \bm{E}$ which indicates that the magnetization induced is parallel to the applied electric field as is schematically shown in Fig.\ref{figure2}b. From  Eq. \ref{ME_isotropic}, it is clear that strong Weyl SOC $v$, long scattering time $\tau$ and large effective mass $m$ can give large magnetoelectric response.

\begin{figure}
\centering
\includegraphics[width=3.6in]{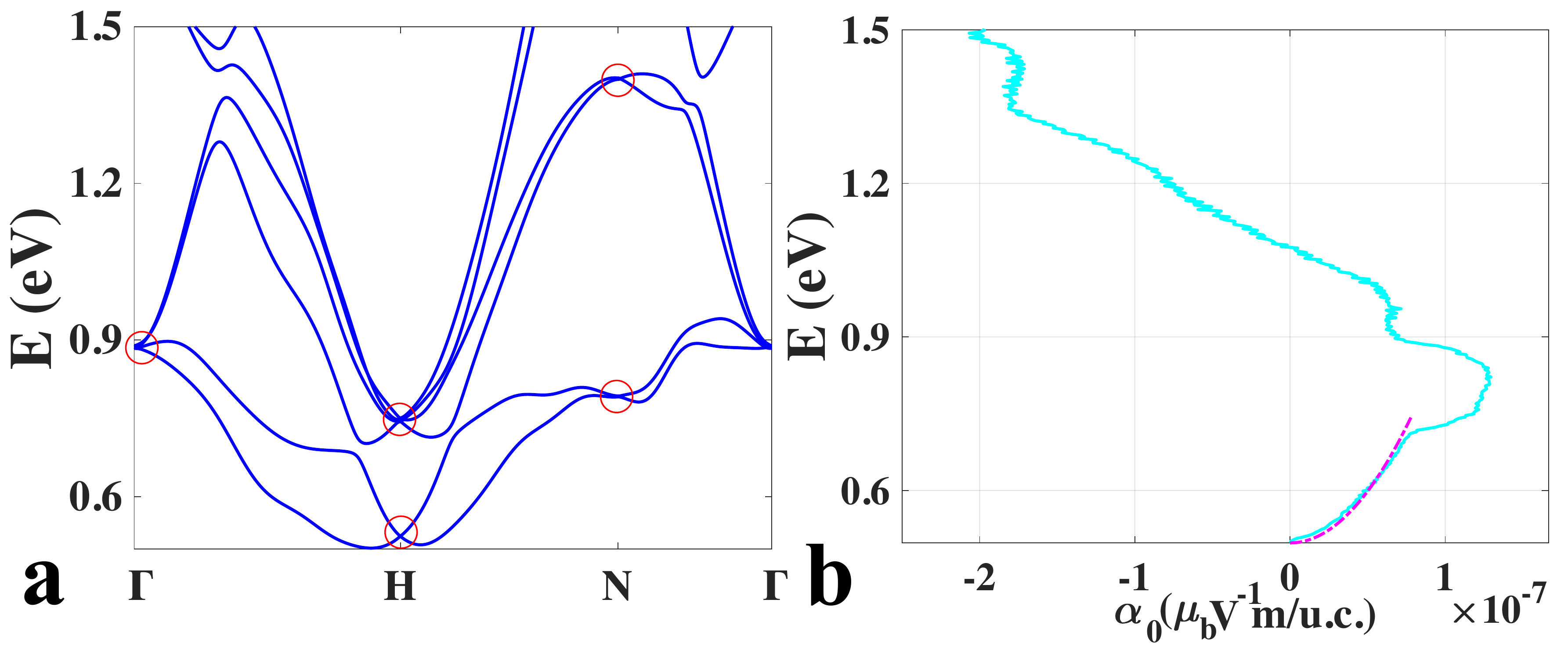}
\caption{The isotropic longitudinal magnetoelectric response in K$_2$Sn$_2$O$_3$. (a) The band structure of K$_2$Sn$_2$O$_3$. The red circles highlight the Kramers Weyl point as well as a multi-fold fermion in this material. (b) The isotropic longitudinal magnetoelectric susceptibility $\alpha_0$ as a function of conduction band energy. Below E=0.7eV, the magnetoelectric response can be well described by the analytical result from Eq. \ref{ME_isotropic} with $m=1.4m_{\textrm{e}}$, $\hbar v=-70$meV$\cdot$nm, $\tau=6$ps. The sudden increase of the magnetoelectric response near E=0.7eV is due to the emergence of the bands associated with the multi-fold fermion at H point of the Brillouin zone. The magenta dashed line denotes the analytical derived $\alpha_0$ in Eq. \ref{ME_isotropic} while the numerically obtained $\alpha_0$ is plotted in a cyan solid line.}\label{figure3}
\end{figure}

To acquire a strong magnetoelectric response, it is also preferable to have long scattering time as shown in Eq.\ref{ME_isotropic}. In KWS, since the Kramers Weyl points are pinned at the time reversal invariant momenta, the electrons on opposite sides of the Weyl point have opposite spin. As a result, for elastic back scattering from scalar impurities, the intra Weyl pocket scattering is suppressed by the SOC, similar to the case in the surface states of topological insulators~\cite{Yazdani, Xue, Shoucheng2}. For an ideal KWS that only has Fermi surfaces enclosing the Weyl point at time reversal invariant momenta, the Weyl SOC can enhance the scattering time by a factor of 3 as shown in the method section using Born approximation. Therefore, we choose a scattering time $\tau=1$ps, which is in the range of the inter Weyl point scattering time in TaAs~\cite{Mertig2, Ramshaw, ShuangJia}. With the effective mass $m=1.4m_e$, Weyl SOC strength $\hbar v=-70$meV$\cdot$nm, the KWS have the band dispersion shown in Fig. \ref{figure2}c. The isotropic magnetoelectric susceptibility $\alpha_0$ as a function of Fermi energy $E_{\textrm{F}}$ is then evaluated as shown in Fig. \ref{figure2}d. For the Fermi pockets enclosing a single Kramers Weyl point, the isotropic magnetoelectric susceptibility $\alpha_0$ increases with the square root of the Fermi energy $E_{\textrm{F}}$.

To seek a large magnetoelectric response from realistic materials, we study two representative materials in the T point group: the K$_2$Sn$_2$O$_3$~\cite{Guoqing}, the  RhSn~\cite{Burkov}. The K$_2$Sn$_2$O$_3$ is a small gap insulator with a Kramers Weyl point at H near the conduction band bottom. Around the Kramers Weyl point, the conduction bands of K$_2$Sn$_2$O$_3$ have large spin splitting shown in Fig.\ref{figure3}a. In the slight n-doped state, the electrons will occupy the Fermi pockets enclosing the single Kramers Weyl point at H so that the isotropic Weyl Hamiltonian can be effectively described by Eq.\ref{isotropic}, with the effective mass $m=1.4m_{\textrm{e}}$ and SOC strength $\hbar v=-70$meV$\cdot$nm. To validate the calculations using an effective Hamiltonian in the form of Eq.\ref{isotropic}, a realistic tight-binding model for K$_2$Sn$_2$O$_3$ is further constructed (see the method section) to calculate the magnetoelectric susceptibility $\alpha_0$, as is shown in Fig. \ref{figure3}b. Below $E=0.7$eV, only the Weyl bands enclosing the Kramers Weyl point at H are involved so that the numerically calculated magnetoelectric susceptibility matches well with the analytical formula in Eq.\ref{ME_isotropic}. It is interesting to note that there is a four-fold fermion around the energy $E=0.75$ eV. As the Fermi surfaces enclosing multi-fold fermion also contribute to the magnetoelectric susceptibility $\alpha_0$~\cite{Felix}, the four-fold fermion at H point induce a sudden increase of $\alpha_0$ when the chemical potential is above 0.7eV. Assuming that the chemical potential lies at $E=0.6$eV (about 80meV above the Weyl point), an electric field of $10^5$V/m is enough to generate a magnetization of $0.005\mu_{\textrm{b}}$ per unit cell.

\begin{figure}
\centering
\includegraphics[width=3.6in]{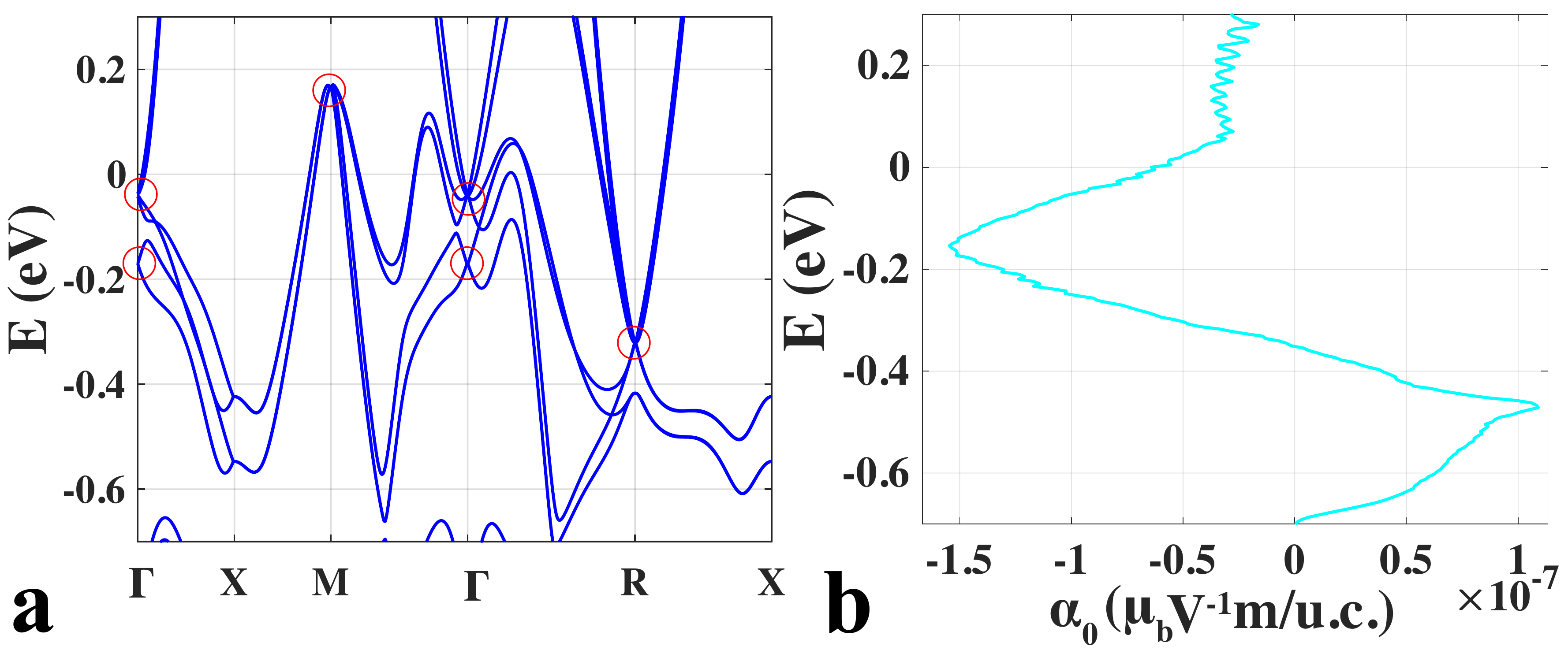}
\caption{The isotropic longitudinal magnetoelectric response in RhSn. (a) The band structure for RhSn. Both Kramers Weyl points as well as other multi-fold fermion band crossing points can be found. The red circles highlight the Kramers Weyl point as well as a four-fold fermion in the mateiral. (b) The isotropic longitudinal magnetoelectric susceptibility $\alpha_0$ in units of $\mu_bV^{-1}\text{m}$ per unit cell is calculated. The effective scattering time is taken to be $\tau=6$ps.}\label{figure4}
\end{figure}

On the other hand, RhSn, which has the same B20 structure as the recently studied KWS RhSi~\cite{Hasan}, is semimetallic. The SOC generates the Kramers Weyl points at $\Gamma$ as well as other multi-fold fermion band crossing points ~\cite{Bradlyn, Shoucheng, Guoqing2} at the time reversal invariant momenta as shown in Fig. \ref{figure4}a.  A four-fold fermion band crossing point near the Fermi energy is highlighted by a red circle in Fig. \ref{figure4}a. With the realistic tight binding model constructed from Wannier functions (see the method section), the magnetoelectric susceptibility $\alpha_0$ is calculated in the energy range including all the bands. The results are shown in Fig. \ref{figure4}b. At the Fermi energy, which is near the multi-fold fermion Weyl point, an electric field of $10^5$V/m can generate a magnetization of 0.007$\mu_{\textrm{b}}$ per unit cell . Compared with the magnetoelectric response of Au(111) surface and Bi/Ag bilayers which have large Rashba SOC~\cite{Mertig1, Mertig2}, the K$_2$Sn$_2$O$_3$ and RhSn can generate magnetizations which are two to three orders of magnitude larger. 

For KWS with dihedral (D$_{n}$) point groups, such as CsCuBr$_3$~\cite{Guoqing}, and elemental Te, Se~\cite{Miyake,Itou, Takeshi, Tetsuaki}, a pure longitudinal magnetoelectric response can also be obtained when the electric field is applied along the direction of the symmetry axes. On the other hand, for KWS with cyclic (C$_{n}$) point groups, such as Ca$_2$B$_5$Os$_3$, a longitudinal magnetoelectric response is generally accompanied with a transverse response. The calculations of magnetoelectric tensors for CsCuBr$_3$ with D$_2$ point group and Ca$_2$B$_5$Os$_3$ with C$_2$ point group can be found in the Supplementary Note 1 and Supplementary Note 2 respectively.

\section{Discussion}
We note that the current induced magnetization in Weyl semimetals was first studied by Johansson et al. in TaAs which belong to the point group C$_{4v}$ with mirror planes~\cite{Mertig2}. Therefore, the Weyl points appear at general $\bm{k}$ points and the resulting spin polarization induced by an electric field is perpendicular to the direction of the electric field. This transverse magnetoelectric effect is similar to the case with Rashba SOC~\cite{Edelstein} which is a property of the polar point group as shown in Supplementary Table 1. If the applied electric field is along the principal axis, however, the magnetoelectric response is zero in TaAs as determined by the C$_{4v}$ point group symmetry.

On the other hand, the longitudinal magnetoelectric response in Weyl semimetals was first studied by Yoda et al.~\cite{Murakami, Yoda}. However, in their models, helical hopping textures~\cite{Murakami, Yoda} are needed for electrons to hop in a spiral manner, imitating the movement of electrons in a classical solenoid. As a result, an orbital magnetization parallel to the direction of an applied electric field would be generated and the longitudinal magnetization is present even without spin-orbit coupling. Unfortunately, no realistic materials that possess such helical hopping textures are identified.

In this work, the finite longitudinal magnetoelecrtric response comes from the current induced net magnetic moments accumulation on Fermi surfaces of the KWS. The net magnetic moments on Fermi surfaces basically have two origins: 1) the SOC induced spin splitting; 2) the finite orbital magnetic moment distribution on Fermi surfaces. The two-fold Kramers Weyl point we focused on in this work contributes to both the spin and orbital magnetic moments, but for a general KWS of chiral crystal symmetry, the complicated band structure diversifies the origin of magnetic moments on Fermi surfaces.

As we have seen, the Fermi pockets enclosing multi-fold fermion contribute to the magnetoelectric susceptibility as well. It is due to the fact that multi-fold fermion endows orbital magnetic moments to the Fermi surfaces outside. Along with the SOC induced spin splitting, the Fermi pockets enclosing multi-fold fermion can also give rise to spin and orbital magnetization driven by charge current~\cite{Felix}. In a KWS, the magnetoelectric response from the multi-fold fermion and two-fold Kramers Weyl point is the same type as the response tensor form is determined by the chiral crystal symmetry of the KWS. Since the Fermi surfaces with a multi-fold fermion inside have different magnetic moment, group velocity and density of states from that with a two-fold Kramers Weyl point, the magnetoelectric susceptibility can differ in the value for the two cases.

Besides the two-fold Kramers Weyl point and multi-fold fermion at high symmetry points in the Brillouin zone, a Weyl point at a general $\bm{k}$ point can generate finite longitudinal magnetoelectric susceptibility as long as the crystal symmetry does not reduce it to zero. This is the case of the local maximal magnetoelectric susceptibility $\alpha_0$ of RhSn at $E=-0.46$eV shown in Fig. \ref{figure4} b. There is a Weyl point in $\Gamma-R$ line around the energy of $E=-0.46$eV which enhances the magnetoelectric susceptibility at that energy. In a brief summary, the longitudinal magnetoelectric susceptibility in KWS of chiral crystal symmetry originates from SOC and Weyl points in the band structure. The magnitude of longitudinal magnetoelectric response at given current density relies on the Fermi surface magnetic moment, Fermi velocity, effective scattering time and density of states at the Fermi energy of KWS.

\begin{figure*}
\centering
\includegraphics[width=\textwidth]{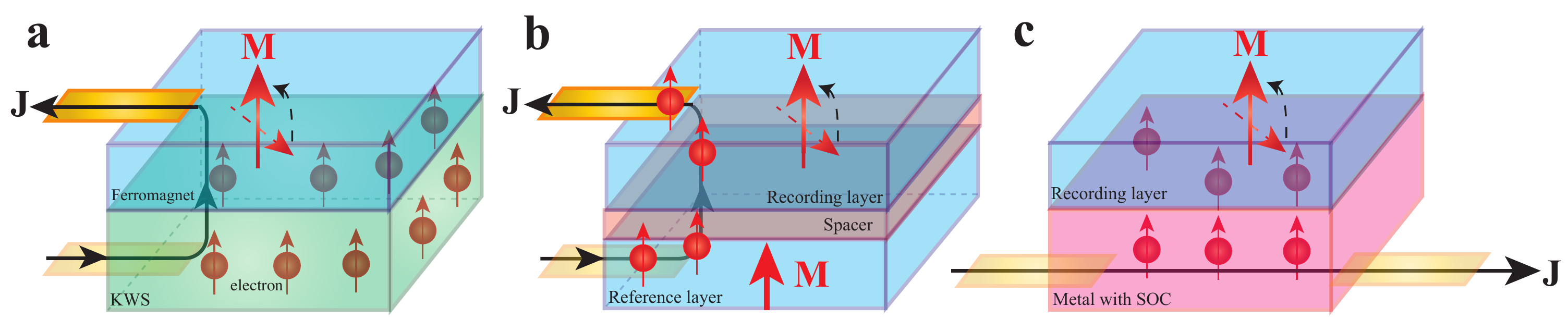}
\caption{The schematics of current induced magnetization switching in spintronic devices. (a) A Kramers Weyl semimetal/ferromagnet heterostructure. When a current $\bm{J}$ passes through the Kramers Weyl semimetal, the effective magnetic field at the Kramers Weyl semimetal and the electrons injected from Kramers Weyl semimetal to the ferromagnetic layer provide a torque to switch the magnetization direction $\left(\bm{M}\right)$ of the ferromagnetic layer. (b) A magnetic tunnelling junction. The junction is made of a reference ferromagnetic layer and the recording ferromagnetic layer separated by a metallic or insulating spacer. The magnetization of the recording layer can be switched by the spin polarized electrons coming out of the reference layer. (c) A spin-orbit torque device~\cite{Manchon, Sinova, Jungwirth, Ralph} . The current induces magnetization at the metal layer through magnetoelectric effect (or the inverse spin galvanic effect). The effective magnetic field at the metal/ferromagnet interface causes the magnetization switching in the ferromagnetic layer.}\label{figure1}
\end{figure*}

Concerning the applications of KWS, due to its unique longitudinal magnetoelectric response, the KWS/ferromagnet heterostructures can be used for new designs of spin-orbit torque devices as shown in Fig.\ref{figure1}a. It is interesting to note that a KWS can cause current induced magnetization switching in the ferromagnetic layer by two effects. First, as depicted in Fig.\ref{figure1}a, the KWS can inject electrons which carry both orbital and spin angular momentum into the ferromagnetic layer. The injection of angular momentum can cause magnetization switching similar to the spin transfer torque induced magnetization switching in magnetic tunnelling junctions as shown in Fig.\ref{figure1}b. The important difference is that the electrons injected by KWS can carry both orbital and spin angular momentum. Particularly, close to the Weyl points, the orbital magnetization carried by the electrons can be significant. Therefore, KWS can work as a source of spin and orbital angular momentum for magnetization switching. Second, when a current is passed along the principal symmetry axis of the KWS, a magnetization is induced at interface between the KWS and the ferromagnet. This current induced magnetization can cause the magnetization switching of the ferromagnet layer through ferromagnetic coupling between the KWS and the ferromagnet.

In recent years, tremendous progress has been made for the study of spin-orbit torque devices as depicted in Fig.\ref{figure1}c. For example, current induced magnetization switching due to spin-orbit torques, in the presence of an in-plane magnetic field, has been realized experimentally in heavy metal/ferromagnet heterostructures~\cite{Gambardella, Mingzhong}. It was also demonstrated that charge currents in multi-layer WTe$_2$ can induce out-of-plane magnetizations~\cite{Ralph} and WTe$_2$ can also be used for magnetization switching for ferromagnets with in-plane anisotropy~\cite{Hyunsoo}. However, an all electric control of magnetization switching for ferromagnets with perpendicular magnetic anisotropy (PMA) through spin-orbit torques, which is important for high density magnetic memories, has not been experimentally realized. Therefore, due to the unique longitudinal response of KWS, they allow new designs of spintronic devices for magnetization switching of ferromagnets with perpendicular magnetic anisotropy as depicted in Fig.\ref{figure1}a.

Concerning the candidate materials of KWS, the cubic chiral B20 structured materials~\cite{Burkov}, such as the recent experimentally studied RhSi~\cite{Hasan}, CoSi~\cite{Hasan, DingHong, Ando}, AlPt~\cite{Yulin}, PtGa~\cite{Felser1}, PdGa~\cite{Felser2}, and RhGe, RhSn all belong to the T point group that has the isotropic longitudinal magnetoelectric response. Particularly, the PtGa~\cite{Felser1} that has the strongest SOC reported so far in chiral crystals is expected to produce strong longitudinal magnetoelectric effect. Besides, elemental Se and Te also have chiral lattice structure~\cite{Miyake, Itou}. Interestingly, a few superconducting materials with chiral lattice symmetry and strong spin-orbit coupling such as Li$_2$Pt$_3$B~\cite{Salamon}, Li$_2$Pd$_3$B~\cite{Hirata}, Mo$_3$Al$_2$C~\cite{Prozorov}, TaRh$_2$B$_2$ and NbRh$_2$B$_2$~\cite{Cava} have been experimentally studied. In their normal state, these superconducting materials can also have spintronic applications. %These materials will also provide a platform to study the interplay between superconductivity, chiral lattice symmetry and magnetoelectric effects.

In conclusion, the magnetoelectric effect of KWS are studied in this work using three different methods: 1. Group theory analysis; 2. Analytical calculations through linear response theory; and 3. Realistic tight-binding calculations are constructed through first principle calculations and the magnetoelectric response of realistic materials are calculated. A new design for current-controlled magnet memory device is proposed and new candidates of KWS are suggested.

\section{Methods}
\subsection{Symmetry Analysis for the Magnetoelectric Response in Chiral Crystals}

In the chiral crystals with magnetoelectric effect $M_i=\sum_{i, j}\alpha_{ij}E_j$ with $i, j= x, y, z$, under the crystal symmetry the magnetization transforms as $M_i\rightarrow\det\left(\hat{R}\right)\hat{R}_{ij}M_j$ while the electric field transforms as $E_i\rightarrow\hat{R}_{ij}E_j$, where $\hat{R}$ represents the symmetry transformation operator and is an orthogonal matrix. As a result, the magnetoelectric susceptibility $\alpha_{ij}$ under the crystal symmetry respects
\begin{align}
\alpha=\det\left(\hat{R}\right)\hat{R}\alpha\hat{R}^{\textrm{T}}.
\end{align}
The chiral point groups, which do not allow improper rotations, can be divided into three sub-classes: the cubic point groups $\left\{\textrm{T}, \textrm{O}\right\}$, the dihedral point groups $\left\{\textrm{D}_n\right\}$ with $n=2, 3, 4, 6$ and the cyclic point groups C$_n$ with $n=1, 2, 3, 4, 6$. In the cubic point groups $\left\{\textrm{T}, \textrm{O}\right\}$, the multiple high order rotation axes along different directions would force the magnetoelectric susceptibility to be proportional to the identity matrix. In the dihedral point groups $\left\{\textrm{D}_n\right\}$ with $n=2, 3, 4, 6$, the C$_n$ rotation axis along $z$ and the in-plane C$_2$ rotation axis along the $x$-axis would eliminate all the off diagonal elements and leave only the diagonal elements in the magnetoelectric susceptibility $\alpha=\textrm{diag}\left\{\alpha_{xx}, \alpha_{yy}, \alpha_{zz}\right\}$. In $\left\{\textrm{D}_3, \textrm{D}_4, \textrm{D}_6\right\}$, the principal axis would further make $\alpha_{xx}=\alpha_{yy}$. In the cyclic point groups $\left\{\textrm{C}_n\right\}$ with $n=1, 2, 3, 4, 6$, the lower symmetry would allow the off-diagonal elements to coexist with the diagonal elements. The explicit forms of the magnetoelectric susceptibility pseudotensor $\alpha_{ij}$ from the symmetry analysis is shown in Table \ref{magnetoelectric_pseudotensor} for the 11 chiral point groups.

\subsection{Magnetoelectric susceptibility in isotropic KWS}
In the isotropic KWS, at the Fermi energy $E_{\textrm{F}}=\frac{\hbar^2\bm{k}^2}{2m}\pm\hbar v|\bm{k}|$, there are two spherical Fermi surfaces with the corresponding wave vectors $\bm{k}_{\textrm{F}\pm}=k_{\textrm{F}\pm}\hat{\bm{k}}$ and $k_{\textrm{F}}=\frac{1}{\hbar}\left(\sqrt{2mE_{\textrm{F}}+m^2v^2}\mp mv\right)$, $\hat{\bm{k}}=\frac{\bm{k}}{|\bm{k}|}$. On the Fermi surfaces enclosing the Kramers Weyl point, the spin magnetic dipole moment and orbital magnetic dipole moment are respectively
\begin{align}
\bm{S}_{\bm{k}_{\textrm{F}\pm}}=\pm\frac{g}{2}\frac{e\hbar}{2m_{\textrm{e}}}\hat{\bm{k}}_{\textrm{F}},\quad\quad\quad\bm{m}_{\bm{k}_{\textrm{F}}\pm}=\pm\frac{1}{2}ev\frac{\hat{\bm{k}}_{\textrm{F}}}{|\bm{k}_{\textrm{F}}|},
\end{align}
with $\pm$ denoting the two band branches. The density of states can be obtained as
\begin{align}
N_{\pm}\left(E_{\textrm{F}}\right)=\frac{m}{2\pi^2\hbar^3}\left(\sqrt{2mE_{\textrm{F}}+m^2v^2}+\frac{mv^2}{\sqrt{\frac{2E_{\textrm{F}}}{m}+v^2}}\mp2mv\right),
\end{align}
and the Fermi velocities there become
\begin{align}
\bm{v}_{\bm{k}_{\textrm{F}\pm}}=\sqrt{\frac{2E_{\textrm{F}}}{m}+v^2}\hat{\bm{k}}_{\textrm{F}\pm}.
\end{align}
At the Fermi energy, the magnetoelectric susceptibility $\alpha_0$ can be approximated through $\sum_n\frac{1}{\left(2\pi\right)^d}\int_{\textrm{BZ}}...\frac{df\left(E_{n, \bm{k}}\right)}{dE_{n, \bm{k}}}d\bm{k}\rightarrow -\sum_nN_n\left(E_{\textrm{F}}\right)\int\frac{d\Omega_{\bm{k}_{\textrm{F},n}}}{4\pi}$ with $n=\pm$ and then we can obtain
\begin{align}
\alpha_0=-\frac{e^2v\tau}{6\hbar^2\pi^2}\sqrt{2mE_{\textrm{F}}+m^2v^2}\left(g\frac{m}{m_{\textrm{e}}}-1\right).
\end{align}

\subsection{Isotropic Weyl SOC Suppressed Back Scattering}
In the pure scalar potential scattering process
\begin{align}
\mathcal{H}_{\textrm{imp}}=\int d\bm{k}d\bm{k}'\psi^\dagger_{\bm{k}'}V_{\bm{k}-\bm{k}'}\psi_{\bm{k}},
\end{align}
the scattering potential can be projected onto the band basis as
\begin{align}
\hat{V}_{\bm{k}-\bm{k}'}=V_{\bm{k}-\bm{k}'}U^\dagger_{\bm{k}'}U_{\bm{k}},
\end{align}
with
\begin{align}
U_{\bm{k}}=\begin{pmatrix}
\cos\frac{\theta}{2}e^{-i\phi} & \sin\frac{\theta}{2}e^{-i\phi}\\ 
\sin\frac{\theta}{2} & -\cos\frac{\theta}{2}
\end{pmatrix},
\end{align}
and $\theta$, $\phi$ are the azimuth and polar angle of $\bm{k}$. In the presence of large SOC, the scattering is dominated in each band branch $\left(\pm\right)$ and the effective scattering potential becomes
\begin{align}
\hat{V}_{\bm{k}-\bm{k}',+}=&V_{\bm{k}-\bm{k}'}\left[\cos\frac{\theta'}{2}\cos\frac{\theta}{2}e^{-i\left(\phi'-\phi\right)}+\sin\frac{\theta'}{2}\sin\frac{\theta}{2}\right]\\
\hat{V}_{\bm{k}-\bm{k}',-}=&V_{\bm{k}-\bm{k}'}\left[\sin\frac{\theta'}{2}\sin\frac{\theta}{2}e^{i\left(\phi'-\phi\right)}+\cos\frac{\theta'}{2}\cos\frac{\theta}{2}\right].
\end{align}
In the Born approximation, the scattering in each band branch is obtained as
\begin{align}
\frac{1}{\tau_\pm}=\frac{mk_{\textrm{F}\pm}n_{\textrm{imp}}}{4\pi^2\hbar^3}\int_0^\pi d\vartheta\int_0^{2\pi} d\varphi\hat{V}_{\bm{k}-\bm{k}', \pm}^2\left(1-\cos\vartheta\right)\sin\vartheta,
\end{align}
with $\varphi=\phi'-\phi$, $\vartheta=\theta'-\theta$. In the spherical Fermi surfaces, the scattering potential $\hat{V}^2_{\bm{k}-\bm{k}', \pm}$ only depends on the angle between $\bm{k}$ and $\bm{k}'$, so
\begin{align}
\int_0^\pi d\vartheta\int_0^{2\pi}d\varphi\hat{V}_{\bm{k}-\bm{k}', \pm}^2\left(1-\cos\vartheta\right)\sin\vartheta=&\frac{172}{105}k_{\textrm{F}\pm}^2,\\
\int_0^\pi d\vartheta\int_0^{2\pi}d\varphi V_{\bm{k}-\bm{k}', \pm}^2\left(1-\cos\vartheta\right)\sin\vartheta=&\frac{16}{3}k_{\textrm{F}}^2.
\end{align}
As a result, the Weyl SOC reduces the scattering potential to about 0.3 and suppresses the back scattering.

\subsection{First-principles Calculation and Wannier Function Construction}
We perform first-principles calculation within the density functional theory framework as implemented in Vienna Abinitio Simulation Package(VASP)~\cite{Kresse}. PAW~\cite{Blochl, Joubert} type of pseudopotential with PBE exchange functional~\cite{Ernzerhof} is used in the calculations and spin orbit coupling is included in the pseudopotentials. After the first-principle calculation is done, tight binding Hamiltonian which perfectly recovers bands near Fermi surface is built through Wannier function construction using package Wannier90~\cite{Vanderbilt, Mostofi}. The lattice structures of all the materials we considered are obtained from the ICSD~\cite{ICSD}.

\section*{Data Availability}
The data generated from our codes that support the findings of this study are available from the corresponding author upon reasonable request.

\section*{Author Contributions}

K. T. L. and W.-Y. H. conceived the idea and initiated the project. W.-Y. H. performed the symmetry analysis and the theoretical calculation of the magnetoelectric susceptibility. X. Y. X. performed the first principles calculation and Wannier function construction for the selected Kramers Weyl semimetals in this work. All the authors discussed the results and co-wrote the paper.

\section*{Acknowledgement}
The authors thank Xi Dai,  Kin Fai Mak, Qiming Shao and Binghai Yan for inspiring discussions. K. T. Law is thankful for the support of HKRGC through C6026-16W, 16324216, 16307117 and 16309718. K. T. Law is further supported by the Croucher Foundation and the Dr. Tai-chin Lo Foundation.
%\bibliography{main}

\section*{Competing Interests}
The authors declare no competing interests.

\onecolumngrid
\clearpage
\begin{center}

{\bf Supplementary Material for ``Kramers Weyl Semimetals as Quantum Solenoids and Their Applications in Spin-Orbit Torque Devices"}
\end{center}

\maketitle
\setcounter{equation}{0}
\setcounter{figure}{0}
\setcounter{table}{0}
\setcounter{page}{1}
\makeatletter
\renewcommand{\theequation}{S\arabic{equation}}
\renewcommand{\thefigure}{S\arabic{figure}}
\renewcommand{\thetable}{S\arabic{table}}
\renewcommand{\bibnumfmt}[1]{[S#1]}
\renewcommand{\citenumfont}[1]{S#1}

\newcommand{\tabincell}[2]{\begin{tabular}{@{}#1@{}}#2\end{tabular}}
\begin{table}[ht]
\caption{List of Magnetoelectric suceptibility pseudotensor $\alpha_{ij}$ and the SOC pseudotensor $v_{ij}$ for crystals in the 18 gyrotropic point groups. Under the crystal symmetry transformation, the $\alpha_{ij}$ and $v_{ij}$ have the same behavior. The elements in $\alpha_{ij}$ and $v_{ij}$ is no longer independent. In T and O point group, $\alpha_{ij}=\alpha_0\delta_{ij}$, $v_{ij}=v\delta_{ij}$. In the group $\left\{\textrm{C}_3, \textrm{C}_4, \textrm{C}_6, \textrm{D}_3, \textrm{D}_4, \textrm{D}_6\right\}$, $\alpha_{\parallel}$ and $v_{\parallel}$ is used to denote $\alpha_{xx}=\alpha_{yy}=\alpha_{\parallel}$, $v_{xx}=v_{yy}=v_{\parallel}$ respectively. The superscript d and - are used to denote the symmetric and anti-symmetric element respectively. The coordinate is set to have principal axis along $z$. $\delta_{ij}$ is the Kronecker delta function with $i, j=x, y, z$.} % title of Table
\centering % used for centering table
\begin{tabular}{c c c c c c} % centered columns (4 columns)
\hline\hline %inserts double horizontal lines\
Point group & $v_{ij}$ & $\alpha_{ij}$ & Point group & $v_{ij}$ & $\alpha_{ij}$\\ [0.5ex] % inserts table
%heading
\hline % inserts single horizontal line
C$_1$ & $\begin{pmatrix}
v_{xx} & v_{xy} & v_{xz} \\ 
v_{yx} & v_{yy} & v_{yz} \\ 
v_{zx} & v_{zy} & v_{zz}
\end{pmatrix}$ & $\begin{pmatrix}
\alpha_{xx} & \alpha_{xy} & \alpha_{xz} \\ 
\alpha_{yx} & \alpha_{yy} & \alpha_{yz} \\ 
\alpha_{zx} & \alpha_{zy} & \alpha_{zz}
\end{pmatrix}$ & C$_{6v}$ & $\begin{pmatrix}
0 & -v^- & 0 \\ 
v^- & 0 & 0 \\ 
0 & 0 & 0
\end{pmatrix}$ & $\begin{pmatrix}
0 & -\alpha^- & 0 \\ 
\alpha^- & 0 & 0 \\ 
0 & 0 & 0
\end{pmatrix}$  \\
C$_2$ & $\begin{pmatrix}
v_{xx} & v_{xy} & 0 \\ 
v_{yx} & v_{yy} & 0 \\ 
0 & 0 & v_{zz}
\end{pmatrix}$ & $\begin{pmatrix}
\alpha_{xx} & \alpha_{xy} & 0 \\ 
\alpha_{yx} & \alpha_{yy} & 0 \\ 
0 & 0 & \alpha_{zz}
\end{pmatrix}$ & D$_{2d}$ & $\begin{pmatrix}
v_{\parallel} & 0 & 0 \\ 
0 & -v_{\parallel} & 0 \\ 
0 & 0 & 0
\end{pmatrix}$ & $\begin{pmatrix}
\alpha_{\parallel} & 0 & 0 \\ 
0 & -\alpha_{\parallel} & 0 \\ 
0 & 0 & 0
\end{pmatrix}$  \\ % inserting body of the table
C$_3$ & $\begin{pmatrix}
v_{\parallel} & -v^- & 0 \\ 
v^- & v_{\parallel} & 0 \\ 
0 & 0 & v_{zz}
\end{pmatrix}$ & $\begin{pmatrix}
\alpha_{\parallel} & -\alpha^- & 0 \\ 
\alpha^- & \alpha_{\parallel} & 0 \\ 
0 & 0 & \alpha_{zz}
\end{pmatrix}$ & S$_{4}$ &  $\begin{pmatrix}
v_{\parallel} & v_{\textrm{d}} & 0 \\ 
v_{\textrm{d}} & -v_{\parallel} & 0 \\ 
0 & 0 & 0
\end{pmatrix}$ & $\begin{pmatrix}
\alpha_{\parallel} & \alpha_{\textrm{d}} & 0 \\ 
\alpha_{\textrm{d}} & -\alpha_{\parallel} & 0 \\ 
0 & 0 & 0
\end{pmatrix}$ \\
C$_4$ & $\begin{pmatrix}
v_{\parallel} & -v^- & 0 \\ 
v^- & v_{\parallel} & 0 \\ 
0 & 0 & v_{zz}
\end{pmatrix}$ & $\begin{pmatrix}
\alpha_{\parallel} & -\alpha^- & 0 \\ 
\alpha^- & \alpha_{\parallel} & 0 \\ 
0 & 0 & \alpha_{zz}
\end{pmatrix}$ & D$_{2}$ & $\begin{pmatrix}
v_{xx} & 0 & 0 \\ 
0 & v_{yy} & 0 \\ 
0 & 0 & v_{zz}
\end{pmatrix}$ & $\begin{pmatrix}
\alpha_{xx} & 0 & 0 \\ 
0 & \alpha_{yy} & 0 \\ 
0 & 0 & \alpha_{zz}
\end{pmatrix}$ \\
C$_6$ & $\begin{pmatrix}
v_{\parallel} & -v^- & 0 \\ 
v^- & v_{\parallel} & 0 \\ 
0 & 0 & v_{zz}
\end{pmatrix}$ & $\begin{pmatrix}
\alpha_{\parallel} & -\alpha^- & 0 \\ 
\alpha^- & \alpha_{\parallel} & 0 \\ 
0 & 0 & T_{zz}
\end{pmatrix}$ & D$_{3}$ & $\begin{pmatrix}
v_{\parallel} & 0 & 0 \\ 
0 & v_{\parallel} & 0 \\ 
0 & 0 & v_{zz}
\end{pmatrix}$ & $\begin{pmatrix}
\alpha_{\parallel} & 0 & 0 \\ 
0 & \alpha_{\parallel} & 0 \\ 
0 & 0 & \alpha_{zz}
\end{pmatrix}$ \\
C$_{1v}$ & $\begin{pmatrix}
0 & v_{xy} & 0 \\ 
v_{yx} & 0 & v_{yz} \\ 
0 & v_{zy} & 0
\end{pmatrix}$ & $\begin{pmatrix}
0 & \alpha_{xy} & 0 \\ 
\alpha_{yx} & 0 & T_{yz} \\ 
0 & \alpha_{zy} & 0
\end{pmatrix}$ & D$_{4}$ & $\begin{pmatrix}
v_{\parallel} & 0 & 0 \\ 
0 & v_{\parallel} & 0 \\ 
0 & 0 & v_{zz}
\end{pmatrix}$  & $\begin{pmatrix}
\alpha_{\parallel} & 0 & 0 \\ 
0 & \alpha_{\parallel} & 0 \\ 
0 & 0 & \alpha_{zz}
\end{pmatrix}$ \\
C$_{2v}$ & $\begin{pmatrix}
0 & v_{xy} & 0 \\ 
v_{yx} & 0 & 0 \\ 
0 & 0 & 0
\end{pmatrix}$ & $\begin{pmatrix}
0 & \alpha_{xy} & 0 \\ 
\alpha_{yx} & 0 & 0 \\ 
0 & 0 & 0
\end{pmatrix}$ & D$_{6}$ & $\begin{pmatrix}
v_{\parallel} & 0 & 0 \\ 
0 & v_{\parallel} & 0 \\ 
0 & 0 & v_{zz}
\end{pmatrix}$ & $\begin{pmatrix}
\alpha_{\parallel} & 0 & 0 \\ 
0 & \alpha_{\parallel} & 0 \\ 
0 & 0 & \alpha_{zz}
\end{pmatrix}$ \\
C$_{3v}$ & $\begin{pmatrix}
0 & -v^- & 0 \\ 
v^- & 0 & 0 \\ 
0 & 0 & 0
\end{pmatrix}$ & $\begin{pmatrix}
0 & -\alpha^- & 0 \\ 
\alpha^- & 0 & 0 \\ 
0 & 0 & 0
\end{pmatrix}$ & T & $\begin{pmatrix}
v & 0 & 0 \\ 
0 & v & 0 \\ 
0 & 0 & v
\end{pmatrix}$ & $\begin{pmatrix}
\alpha_0 & 0 & 0 \\ 
0 & \alpha_0 & 0 \\ 
0 & 0 & \alpha_0
\end{pmatrix}$\\
C$_{4v}$ & $\begin{pmatrix}
0 & -v^- & 0 \\ 
v^- & 0 & 0 \\ 
0 & 0 & 0
\end{pmatrix}$ & $\begin{pmatrix}
0 & -\alpha^- & 0 \\ 
\alpha^- & 0 & 0 \\ 
0 & 0 & 0
\end{pmatrix}$ & O & $\begin{pmatrix}
v & 0 & 0 \\ 
0 & v & 0 \\ 
0 & 0 & v
\end{pmatrix}$ & $\begin{pmatrix}
\alpha_0 & 0 & 0 \\ 
0 & \alpha_0 & 0 \\ 
0 & 0 & \alpha_0
\end{pmatrix}$ \\[1ex] % [1ex] adds vertical space
\hline %inserts single line
\end{tabular}
\label{magnetoelectric_pseudo-tensor} % is used to refer this table in the text
\end{table}

\section{Suppelementary Note 1: Anisotropic Longitudinal Magnetoelectric Response in KWS in Dihedral Point Groups}
In the KWS with dihedral point group symmetry, the magnetoelectric susceptibility has the nonzero value only in the diagonal elements. This indicates that the magnetization generated is parallel to the electric field if the electric field is applied along the symmetry axes. For example, the KWS CsCuBr$_3$ belongs to the D$_2$ point group and its realistic band structure in Supplementary Figure \ref{figureS1}a shows the Kramers Weyl points at the time reversal invariant momenta \{$\Gamma$, Y, S, Z\}. Near the Kramers Weyl points, the bands splittings are highly anisotropic as allowed by the D$_2$ point group symmetry. The anisotropic longitudinal magnetoelectric susceptibility $\left\{\alpha_{xx}, \alpha_{yy}, \alpha_{zz}\right\}$, as a function of the Fermi energy, are numerically evaluated based on the realistic band structure, are depicted in Supplementary Figure \ref{figureS1}b. 

\section{Suppelementary Note 2: Coexistence of Longitudinal and Transverse Magnetoelectric response in KWS in Cyclic Point Groups}
The KWS in the cyclic point group possess a polar axis that leaves the crystal invariant under rotation. This symmetry allows both the longitudinal and the transverse magnetoelectric response: an electric field along the polar axis can generate pure longitudinal magnetization while the electric field inside the basal plane produces magnetization with nonzero parallel and perpendicular components. In Ca$_2$B$_5$Os$_3$ with a single C$_2$ rotation axis, the spin splitted bands with Kramers Weyl points are shown in Supplementary Figure \ref{figureS1}c. Based on the realistic band structure, the nonvanishing magnetoelectric susceptibility elements under the C$_2$ symmetry are numerically calculated at different energy. The results are shown in Supplementary Figure \ref{figureS1}d and large magnetoelectric response is also obtained due to the strong spin-orbit coupling of the material.

\begin{figure}
\centering
\includegraphics[width=\textwidth]{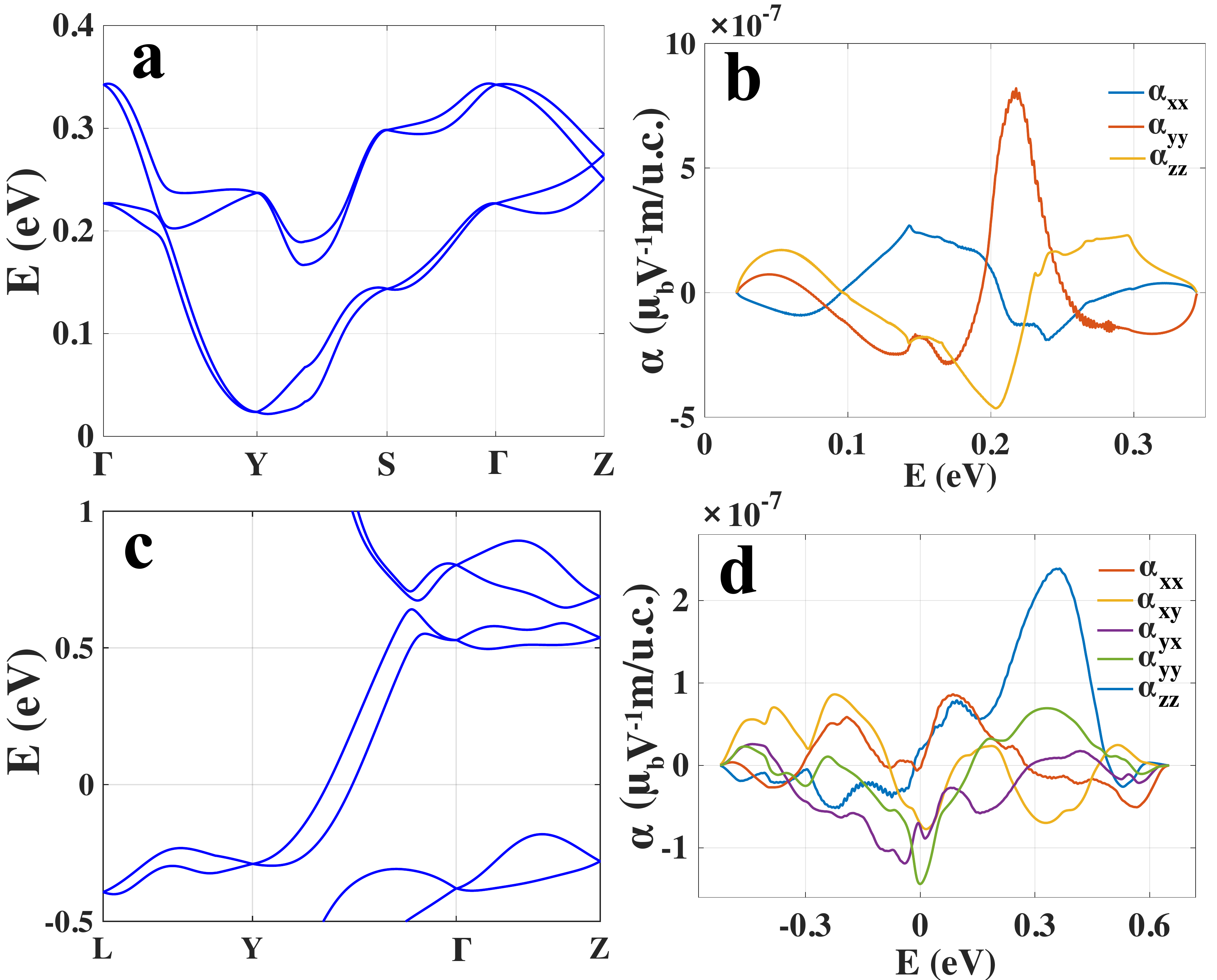}
\caption{The magnetoelectric effect in CsCuBr$_3$ and Ca$_2$B$_5$Os$_3$. (a) The band structure for CsCuBr$_3$. (b) The anisotropic longitudinal magnetoelectric susceptibility $\left\{\alpha_{xx}, \alpha_{yy}, \alpha_{zz}\right\}$ as a function of the band energy. (c) The band structure for Ca$_2$B$_5$Os$_3$. (d) The nonvanishing magnetoelectric susceptibility elements as a function of the band energy. The effective scattering time is taken to be $\tau=6$ps. u.c. is short for unit cell.}\label{figureS1}
\end{figure}

\end{document}